\documentclass[a4paper,11pt]{article}
\pdfoutput=1 

\usepackage{jheppub} 
\usepackage{amsmath}
\usepackage{slashed}
\usepackage{amssymb}
\usepackage[mathscr]{euscript}
\usepackage{enumerate}
\usepackage{hyperref}
\usepackage{subfigure}

\newcommand{\hs}{\hspace{0.15mm}}

\newcommand{\ms}{\mathbb{S}}

\title{Asymptotic Symmetries from finite boxes}

\author[a]{Tom\'as Andrade}
\author[b]{Donald Marolf}

\affiliation[a]{Rudolf Peierls Centre for Theoretical Physics \\ University of Oxford, 1 Keble Road, Oxford OX1 3NP, UK}
\affiliation[b]{University of California at Santa Barbara \\ Santa Barbara, CA 93106, USA}

\emailAdd{tomas.andrade@physics.ox.ac.uk}
\emailAdd{marolf@physics.ucsb.edu}

\abstract{It is natural to regulate an infinite-sized system by imposing a boundary condition at finite distance, placing the system in a ``box."  This breaks symmetries, though the breaking is small when the box is large.  One should thus be able to obtain the asymptotic symmetries of the infinite system by studying regulated systems.  We provide concrete examples in the context of Einstein-Hilbert gravity (with negative or zero cosmological constant) by showing in 4 or more dimensions how the Anti-de Sitter and Poincar\'e asymptotic symmetries can be extracted from gravity in a spherical box with Dirichlet boundary conditions.  In 2+1 dimensions we obtain the full double-Virasoro algebra of asymptotic symmetries for AdS$_3$ and, correspondingly, the full Bondi-Metzner-Sachs (BMS) algebra for asymptotically flat space.  In higher dimensions, a related approach may continue to be useful for constructing a good asymptotically flat phase space with BMS asymptotic symmetries.}

\begin{document}
\maketitle
\flushbottom

\section{Introduction} \label{sec:intro}

Asymptotic symmetries are a central tool in the study of infinite gravitational systems.  Nevertheless,  to the uninitiated they often present both conceptual and computational challenges.  Both sets of issues arise because they represent diffeomorphisms that cannot be considered pure gauge due to often-subtle details of the boundary conditions.  In particular, when confronted with an infinite system it is not always apparent precisely which choice of boundary conditions will lead to physically interesting symmetries.  The ensuing cycles of trial and error can then absorb much effort.  This is exemplified by the study of asymptotically flat spacetimes in $3+1$ dimensions, where despite much history (see e.g. \cite{Arnowitt:1962hi,Bondi:1962px,Sachs:1962zza,Regge:1974zd,Ashtekar:1978zz,Dray:1984rfa,Ashtekar:1981bq,Ashtekar:1990gc}) , recent physical arguments \cite{Strominger:2013jfa,Kapec:2014opa,Strominger:2014pwa,Pasterski:2015tva,Kapec:2015vwa},   and creative attempts \cite{Compere:2011ve}, there is no known construction of a phase space with on which the Bondi-Metzner-Sachs (BMS) group \cite{Bondi:1962px,Sachs:1962zza} acts as an asymptotic symmetry and for which the symplectic structure is conserved between past and future null infinity.

On the other hand, it is natural to regulate infinite systems by imposing boundary conditions at finite distance, often described as placing the system in a box.  This idea has a long history in the gravitational context (see e.g. \cite{Hawking:1976de,Hawking:1982dh,York:1986it,Brown:1986nw,Braden:1990hw,Page:1991kh,Brown:1992br,Brown:1994su,Martinez:1990sd,Brown:1990fk,Brady:1991np,Louko:1994tv,Fischetti:2012rd}) where it is common to impose a Dirichlet boundary condition, fixing the induced metric at the walls of the box\footnote{Though see \cite{2006math.....12647A} for an interesting alternative.}.  The construction is quite concrete, and the physical nature of certain diffeomorphisms is clear: a diffeomorphism that changes the relationship of bulk objects to points on the boundary cannot be pure gauge.  We will in particular focus below on diffeomorphisms that change the distance between bulk objects (stars, planets, black holes...) and points on the boundary.  Such excitations are naturally interpreted as finite displacements of the bulk center-of-mass.

When the box is much larger than its contents, one expects the regulated system to admit an approximate notion of asymptotic symmetries.  Here we have in mind some well-defined transformation of the finite system with finite walls -- and in particular which exactly preserves the above-mentioned Dirichlet boundary condition -- but which need not be a symmetry of the regulated system.  This means that it need not preserve the symplectic structure of the phase space, and so need not be generated by the Poisson Bracket with some observable.  In other words, while the transformation can be thought of as some flow on the phase space, and is thus generated by some phase space vector field, the vector field need not be Hamiltonian.  Instead, it is merely the assignment of a linearized deformation to each solution.  It is only in an appropriate infinite-volume limit where the regulator is removed that it becomes an asymptotic symmetry.

Our purpose here is to demonstrate in simple examples how the asymptotic symmetries of infinite systems can be recognized in regulated systems with finite-distance walls having fixed induced metric, henceforth referred to as Dirichlet walls.  The work is exploratory; we do not attempt a full construction of the infinite volume phase as a limit of such Dirichlet wall systems.  We concentrate on transformations whose action on a given solution takes the form of a diffeomorphism.  It may thus be thought of as defined by a spacetime vector field on each solution, though we allow this vector field to depend on the solution in an arbitrary way.  In practice, we simply seek linearized diffeomorphisms about given solutions that preserve the desired boundary conditions and which define non-trivial directions of the symplectic structure.  In the limit where the system becomes infinite these are induced by vector fields that become independent of the solution, indicating the emergence of an asymptotic symmetry.

Though a few new calculations are required, our task largely consists of assembling results from the literature.  For black holes inside spherical Dirichlet walls in $d\ge 4$ spacetime dimensions, \cite{Andrade:2015gja} recently identified linearized diffeomorphisms with dipole ($j=1$) profiles that describe physical motion of the black holes away from the center of the box.  We simply note in section \ref{kick} that they become independent of black hole parameters in the large-box limit, and that their symplectic products reproduce\footnote{\label{HamVec} Recall the basic rule of classical mechanics that the Poisson bracket of two observables is the symplectic product of associated Hamiltonian vector fields; i.e., of the transformations they generate on phase space. For general observables $A,B$, phase space coordinates $\xi$, and symplectic product $\Omega$ we may write $\{A, B \} = \Omega(\delta_A \xi, \delta_B \xi)$. } the Poincar\'e or anti-de Sitter (AdS) algebra for respectively zero or negative cosmological constant ($\Lambda$).  Though it may also be interesting, we do not study the case of positive cosmological constant as the finite-sized cosmological horizon makes any large-box limit physically very different.  We also show for all cases that $j>1$ diffeomorphisms preserve the boundary conditions only when they vanish at the wall and so are pure gauge.

For $d=3$ and $\Lambda <0$ it is also known (see section 5 of \cite{Marolf:2012dr}) that BTZ black holes inside Dirichlet walls admit a large family of linearized diffeomorphisms preserving boundary conditions but changing the distance between the horizon and various points on the boundary.  Indeed, at a given time they correspond to displacing the rotationally-symmetric boundary to an arbitrary new surface outside the black hole.  In the large box limit, such diffeomorphisms clearly become the usual AdS$_3$ boundary gravitons associated with the asymptotic Virasoro algebras. Section \ref{excBTZ} studies these diffeomorphims in a mode decomposition and computes their symplectic products to explicit the recovery of the usual double-Virasoro algebra.  It is then straightforward to follow \cite{Barnich:2006av} and take $\Lambda \rightarrow 0$ to recover the 2+1 Bondi-Metzner-Sachs (BMS) algebra for the asymptotically flat case.  The limits commute, so one may also work directly with $\Lambda =0$ to recover 2+1 BMS from $\Lambda =0$ systems with finite Dirichlet walls.  We close with brief comments on future directions in section \ref{disc}.  Discussion of the symplectic structure in the presence of Dirichlet walls is relegated to appendix \ref{SSB}.

\section{Kicking a Schwarzschild(-AdS) black hole}
\label{kick}

Recall that \cite{Andrade:2015gja} studied
perturbations of Schwarzschild and Schwarzschild-AdS black holes in $d \ge 4$ dimensions with metric
\begin{equation}\label{Schw background}
	ds^2 = - f(r) dt^2 + \frac{dr^2}{f(r)} + r^2 d \sigma_n^2 \ \ \ \text{for} \ \ \  \qquad f(r) =  \frac{r^2}{\ell^2}  + 1 - \frac{2 MG}{r^{n-1}},
\end{equation}
surrounded by a spherical Dirichlet wall at $r=r_D$.
Here $\sigma_{ij}$ is the metric on a $n$-dimensional unit sphere with $n=d-2$ and $M$, $\ell$ are related to the total
energy $E$ and the cosmological constant $\Lambda$ by
\begin{equation}\label{energy}
	E = \frac{n M {\cal A}_n}{8 \pi}, \qquad \Lambda = -\frac{n(n+1)}{2\ell^2}, 	
\end{equation}
\noindent where ${\cal A}_n = \frac{2 \pi^{\frac{n+1}{2}}}{\Gamma\left({\frac{n+1}{2}}\right)}$ is the area of the unit $n$-sphere.
In particular, it was found that linearized diffeomorphisms preserve the Dirichlet boundary conditions when they were
generated by vector fields that (with indices lowered by \eqref{Schw background}) satisfy
\begin{equation}
\label{Svec}
\xi_t = e^{- i \omega t} c_t \mathbb{S}, \ \ \
\xi_r = e^{- i \omega t} c_r \mathbb{S}, \ \ \
\xi_i = - e^{- i \omega t} \frac{r}{\sqrt{n}}  L(r) {\cal D}_i \mathbb{S},
\end{equation}
for
\begin{align}
\label{relations}
\nonumber
c_t(r_D) &= -	\frac{i \omega}{\sqrt{n}} r_D L(r_D), \\
c_r(r_D) &= - \frac{ L(r_D)}{\sqrt{n} f(r_D)}, \\
\nonumber
\omega & = \pm \sqrt{\frac{f'(r_D)}{2r_D}},
\end{align}
and any function $L(r)$.  Here ${\cal D}_i$ is the covariant derivative on the unit $S^n$ and $\mathbb{S}$ is a scalar spherical harmonic with angular momentum $j=1$.  the functions $c_r,c_t$ are unconstrained away from $r=r_D$. Note that $f'(r_D)$ is positive for \eqref{Schw background} so our frequencies are real. For $d=4$, \cite{Andrade:2015gja} also computed symplectic products to check that such linearized diffeomorphisms represent physical disturbances -- i.e., that they are not pure gauge -- when both $M$ and $L(r_D)$ are non-zero.  However, they do become pure gauge when there is no bulk object to displace relative to the wall ($M=0$) and when the diffeomorphism acts trivially at the wall ($L(r_D) =0$).  On general grounds (see e.g. \cite{Brown:1992br,Hollands:2005ya,Fischetti:2012rd}), diffeomorphisms which induce isometries of the boundary define exact symmetries of the Dirichlet wall system generated by non-trivial charges.  But we instead focus on symmetries broken by our regulator, corresponding to diffeomorphisms that displace the wall as in \eqref{Svec}.

In analogy with the Klein-Gordon inner product for scalars fields, for oscillatory modes it is useful to define the inner product
\begin{equation}
\label{IP}
	(\delta_1 g_I, \delta_2 g_J) = - i \Omega(\delta_1 g_I, \delta_2 g^*_J),
\end{equation}
\noindent where $^*$ denotes complex conjugation and $\Omega(\delta_1 g, \delta_2 g^*)$ is the symplectic product of linearized metrics $\delta_1 g_I, \delta_2 g_J$ generated by the above vector fields with associated functions $L_1(r), L_2(r)$ and spherical harmonics $\mathbb{S}_I, \mathbb{S}_J$.  Generalizing the calculation of \cite{Andrade:2015gja} yields
\begin{equation}\label{ip l=1}
	(\delta_1 g_I, \delta_2 g_J) = \frac{4 (d-1) MG \omega}{16 \pi G f(r_D)}   L_1(r_D) L_2^*(r_D) \delta_{IJ}.
\end{equation}
Here we have chosen the $\ms_{I}$ orthonormal: $\int \sqrt{\sigma} \ms_{I} \ms_{J}^* = \delta_{IJ} $, where $\sqrt{\sigma}$ is the volume element on the unit $S^{d-2}$ and $I$ collectively denotes
all angular quantum numbers.

The important observation is that the relations \eqref{relations} depend on the background (i.e., on the parameter $M$) only through $f, f'$ evaluated at $r_D$.  For large $r_D$ these satisfy $f(r_D) \sim \ell^{-2} r^2_D$, $f'(r_D) \sim \frac{2 r_D}{\ell^2}$ and thus become independent of $M$.  In particular, for finite $\ell$ we take the asymptotic behaviour of $L$ to be
\begin{equation}\label{L asympt Gamma}
	L(r) = \sqrt{\frac{n \pi^{(d-1)/2}}{2\Gamma(\frac{d+1}{2})}} r \ell + O(1),
\end{equation}
\noindent so in the large $r$ limit we find
\begin{align}\label{AdSg}
(\delta_1 g_I, \delta_2 g_J) & \rightarrow \ell E \delta_{IJ}.
\end{align}
Here $E$ is the total
energy of the black hole given by \eqref{energy}. We note that the numerical factor in \eqref{L asympt Gamma} follows from the normalization
condition on the spherical harmonics.

We now explain how \eqref{AdSg} gives the AdS algebra of asymptotic symmetries.
First, we observe that the harmonic time dependence in \eqref{Svec} means that the diffeomorphisms are equally well characterized as
pure (positive) displacements in a constant $t$ slice respectively at $t=0$ and $\frac{\pi}{2\omega} \rightarrow \pi \ell/2$, corresponding to their real and imaginary
parts. (At other times they are combinations of such displacements and boosts of the slice.)

Second, it will be convenient to think of AdS$_d$ as the (covering space of the) hyperboloid
\begin{equation}
(T^1)^2 + (T^2)^2 - \sum_{i=1}^{d-1} (X^i)^2 = \ell^2
\end{equation}
in the Minkowski space $M^{d-1,2}$ with signature $(+, \dots, +, -, -)$, coordinates $X^i, T_1, T_2$, and metric
\begin{equation}
ds^2 = \sum_{i=1}^{d-1} (dX^i)^2 - (dT^1)^2 - (dT^2)^2.
\end{equation}
The AdS isometries are then the rotations $J_{X^iX^j}$, the time translation $J_{T^1 T^2}$, and the boosts $K_{T_1 X^i}$, $K_{T_2 X^i}$.  The rotations and time translations are exact symmetries of the regulated Dirichlet wall system, so we focus on the boosts.  The two generators $K_{T_1 X^i}$, $K_{T_2 X^i}$ for the same $i$ are related by a $\pi \ell/2$ time translation and so correspond precisely to the two parts of our diffeomorphism.  Checking the normalizations shows that, for the appropriate linear combinations of spherical harmonics and using the rule described in footnote \ref{HamVec}, one finds that \eqref{AdSg} corresponds to \begin{equation}
[\frac{1}{\sqrt{2}}\left(K_{T_1 X^i} + i K_{T_2 X^i}\right), \frac{1}{\sqrt{2}}\left(K_{T_1 X^i} - i K_{T_2 X^i}\right)] = J_{T^1 T^2} = \ell E,
\end{equation}
as desired. 

To study $\Lambda =0$, we note that the limit $\ell \to \infty$ transforms the real part of the diffeomorphism into $\ell$ times a displacement along $X^i$, and transforms the imaginary part into
along. Equation \eqref{AdSg} thus represents the commutator 
\begin{equation}
[\frac{1}{\sqrt{2}}\left(\ell P_i + i K_{i}\right), \frac{1}{\sqrt{2}}\left(\ell P_i - i K_{i}\right)] = \ell P_0,
\end{equation}
involving the momentum $P_i$, the corresponding boost generator $K_i$, and the energy $P_0$.  One may remove the distracting factors of $\ell$ by writing
\begin{align}\label{AdSg v2}
2i (\text{Re} \left[ \delta_1 g_I \right], \text{Im} \left[ \hat \delta_2 g_J \right] ) & \rightarrow  P_0 \delta_{IJ},
\end{align}
for $\hat \delta_2 g_J  = \frac{1}{\ell} \delta_2 g_J$ and noting that $\text{Re} \left[ \delta_1 g_I \right], \text{Im} \left[ \hat \delta_2 g_J \right]$ both have finite limits as $\ell \rightarrow \infty$.  The remainder of the Poincar\'e algebra involves rotations and further time-translations.   Since these are exact symmetries at finite $r_D$, their commutators -- with the emergent $P_i$ and $K_i$, or with each other -- trivially match those of the Poincar\'e algebra at large $r_D$.

Finally, for $d > 3$ we note that $j > 1$ linearized diffeomorphisms deform the metric on the sphere. They thus violate our Dirichlet boundary conditions unless they vanish at $r_D$.   In the language of \cite{Andrade:2015gja}, which closely follows \cite{Kodama:2003jz}, these diffeomorphisms generate a non-zero component $H_T \propto L$, while the boundary conditions require $H_T =0$ at $r_D$.

\section{Diffeomorphism Excitations of BTZ}
\label{excBTZ}

We now turn to BTZ black holes \cite{Banados:1992wn} surrounded by a Dirichlet wall on which the induced metric is the static cylinder defined by a circle of circumference $2 \pi \ell \rho_D$ in terms of a dimensionless parameter $\rho_D$ and the AdS$_3$ scale $\ell$. In particular, the metric on the wall will be
\begin{equation}
\label{wall}
ds^2_{wall} = - \ell^2( dT^2 + \rho_D^2 d\Phi^2).
\end{equation}

It is useful to begin with the BTZ line element
\begin{equation}
\label{BTZ}
	ds^2 = \tilde f (r)^{-1} dr^2 - \tilde f(r) dt^2 + r^2\left(d \phi - \frac{r_+ r_-}{\ell r^2} dt\right)^2 \ \ \ \text{for} \ \ \ 
	\tilde f(r) = \frac{(r^2 - r_+^2)(r^2 - r_-^2)}{\ell^2 r^2}.
\end{equation}
 The usual mass (energy) and angular momentum of the black hole as measured from infinity are\footnote{Note that our conventions differ from those in \cite{Banados:1992wn} in that we keep track of the factors of
$G$ while they set $8 G = 1$. This is consistent with our normalization for the action in \eqref{total action app}. }
\begin{equation}
	M = \frac{r_+^2 + r_-^2}{8G\ell^2}, \qquad J = \frac{r_+ r_-}{4 G\ell}.
\end{equation}

Introducing dimensionless coordinates $\rho = r/\ell$, $\tau = t/\ell$, the line element reads
\begin{equation}\label{ds2 BTZ}
	ds^2 = \ell^2 \left[ f(\rho)^{-1} d\rho^2 - f(\rho) d\tau^2 + \rho^2\left(d \phi - \frac{\rho_+ \rho_-}{\rho^2} d\tau\right)^2 \right]  \ \  \ \text{for} \ \  \ 
	f(\rho) = \frac{(\rho^2 - \rho_+^2)(\rho^2 - \rho_-^2)}{\rho^2},
\end{equation}
\noindent with $\rho_\pm = r_\pm / \ell$. In such solutions we may take the Dirichlet wall to lie at $\rho = \rho_D$ by defining $T = \sqrt{f}\tau$ and $\Phi = \phi - \frac{\rho_+ \rho_-}{\rho^2}\tau$.

We now seek linearized diffeomorphisms that act non-trivially on the wall while preserving the induced metric \eqref{wall}.   It is convenient to Fourier transform in $\tau, \phi$.  A general vector field  $\xi^\mu = (\xi^\rho, \xi^\tau, \xi^\phi)$ is then a sum of terms which (when the indices are lowered by \eqref{ds2 BTZ}) take the form
\begin{align}\label{xi diff}
	\chi_\mu &= e^{i \omega \tau + i m \phi} ( c_\rho(\rho), c_\tau(\rho),  \rho L(\rho)  ), \\
\label{xi diff 2}
	\bar \chi_\mu &= e^{i \omega \tau - i m \phi} ( c_\rho(\rho), c_\tau(\rho),  - \rho L(\rho)  ).
\end{align}
The field $\chi_\mu$ generates the perturbation
\begin{align}
	\delta g_{\rho \rho} &= 2 c_\rho' + f'/f c_\rho, \\
	\delta g_{\rho \tau} &=  i \omega c_\rho + c_\tau' - \left(\frac{f'}{f} + \frac{2 \rho_+^2 \rho_-^2}{\rho^3 f} \right)
	\left(c_\tau  + \frac{\rho_+ \rho_-}{\rho} L \right), \\
	\delta g_{\rho \phi} &= i m c_\rho  + \rho L' + \left( \frac{2 \rho_+^2 \rho_-^2}{\rho^2 f} - 1\right) L +
	\frac{2 \rho_+ \rho_-}{\rho f} c_\tau,\\
	\delta g_{\tau \tau} &=  2 i \omega c_\tau - f \left(f' + \frac{2 \rho_+^2 \rho_-^2}{\rho^3} \right)  c_\rho, \\
	\delta g_{\tau \phi} &= i ( m c_\tau + \omega \rho L ), \\
	\delta g_{\phi \phi} &= 2 \rho ( f c_\rho + i m L ).
\end{align}
Since $f' + \frac{2 \rho_+^2 \rho_-^2}{\rho^3} = 2\rho$,  to preserve the induced metric at $\rho = \rho_D$ a diffeomorphism with $L(\rho_D) \neq 0$  must satisfy
\begin{equation}\label{cr ct Dbc}
	c_\rho(\rho_D) = - \frac{i mL(\rho_D)}{f(\rho_D)} , \qquad c_\tau(\rho_D) = - \frac{ \omega \rho_D L(\rho_D)}{m},
\end{equation}
\noindent and also
\begin{equation}\label{w for Dbc}
	\omega = m.
\end{equation}
So in agreement with \cite{Marolf:2012dr} we find both purely left-moving and purely right-moving allowed linearized diffeomorphisms given by $\chi^\mu$ and $\bar \chi^\mu$ for each $m$.
Note that the frequency in \eqref{w for Dbc} is completely independent of $\rho_D, \rho_+, \rho_-$.   In fact the entire diffeormophism becomes independent of $\rho_D, \rho_+, \rho_-$ at large $\rho_D$ if we choose
\begin{equation}\label{L asympt}
	L(\rho) = \frac{\rho}{2} \ell^2 + \ldots \ , \ \ \  c_\rho = - \frac{i mL(\rho)}{\rho^4} + \dots \ , \ \ \
c_\tau(\rho_D) = - \frac{ \omega \rho L(\rho)}{m} + \dots .
\end{equation}
For comparison, the usual AdS$_3$ vector fields corresponding to the asymptotic Virasoro symmetries may be taken to be \cite{Brown:1986nw}
\begin{align}\label{asa vec}
	\xi_n &= \frac{i}{2} e^{i n (\tau+\phi)}\left \{  - i n \rho \partial_{\rho}
	+ \left( 1 - \frac{n^2}{2 \rho^2} \right)  \partial_\tau + \left( 1 + \frac{n^2}{2 \rho^2} \right)
	\partial_\phi \right \}, \ \ \ \text{and} \\
\label{asa vec 2}
	\bar \xi_n &= \frac{i}{2} e^{i n (\tau-\phi)}\left \{  - i n \rho \partial_{\rho}
	+ \left( 1 -  \frac{n^2}{2 \rho^2}  \right)  \partial_\tau - \left( 1 -  \frac{n^2}{2 \rho^2} \right)  \partial_\phi \right \}.
\end{align}
Using \eqref{w for Dbc} and \eqref{L asympt}
the generators \eqref{xi diff} agree asymptotically with \eqref{asa vec}, \eqref{asa vec 2} at large $\rho$.

It remains to show that our diffeomorphisms with $L(\rho_D) \neq 0$ define non-trivial excitations; i.e., that they are not pure gauge.  We do so by computing their symplectic products using \eqref{j EH}, \eqref{bndycurrent}, \eqref{Omega full}.  As explained in detail in the appendix, the symplectic product can be written as a bulk integral whose structure is determined by the Einstein-Hilbert Lagrangian, together with a boundary term specific to the case of Dirichlet boundary conditions.  For linearized diffeomorphisms the integrand
in the bulk contribution becomes a total derivative and so depends only on the boundaries; see e.g. (2.8) of \cite{Andrade:2013wsa}.  For the left moving modes $\chi$, taking the only boundary to be at $\partial M$ or taking $\chi$ to vanish near any other boundaries and using \eqref{L asympt}, the analogue of \eqref{IP} yields
\begin{equation}\label{ip rhoD}
	(\delta g_1, \delta g_2) = \frac{1}{2 G} \frac{m_1[8G(M - J/\ell) + m_1^2]}{f(\rho_D) \ell^3} L_1(\rho_D) L_2(\rho_D)^*
	 \delta_{m_1, m_2},
\end{equation}
\noindent where $\delta g_{1,2} = {\cal L}_{\chi} g$ with $\omega = m_1,m_2$.

In particular, the result is conserved because modes with different frequencies are orthogonal.
Inner products of the right-moving diffeomorphisms \eqref{xi diff 2} are obtained by sending $J \to - J$
in \eqref{ip rhoD} . \\

The inner products \eqref{ip rhoD} are non-zero, and simplify at large $\rho_D$ to become
\begin{equation}\label{ip Vir}
	(\delta g_1, \delta g_2) = \frac{\ell}{8G}
	m_1[ 8G(M -  J/\ell) + m_1^2] \delta_{m_1, m_2}.
\end{equation}
As noted in the introduction (see footnote \ref{HamVec} and \eqref{IP}), if the linearized transformations define a Hamiltonian vector field then this also gives the commutator of the relevant generators evaluated on our BTZ background.  Indeed, \eqref{ip Vir} coincides with the left-moving Virasoro algebra
\begin{equation}\label{Vir alg}
[L_{n}, L_m] = (m-n)L_{m+n}	 + \frac{c}{12} (m^3 - m) \delta_{m, -m}
\end{equation}
evaluated on BTZ using the Brown-Henneaux identifications \cite{Brown:1986nw}
\begin{equation}
\label{BH}
c = \frac{3\ell}{2G},  \ \ \ 	L_0 = \frac{1}{2} \left(M\ell - J\right) + \frac{c}{24} = \frac{1}{2} \left( M\ell - J \right) + \frac{\ell}{16 G} ,
\end{equation}
and the fact that $L_m$ vanishes for the solution \eqref{BTZ} when  $m\neq 0$. 

\subsection{The asymptotically flat limit}

We may also study linearized diffeomorphisms of solutions with zero cosmological constant.  This just requires taking the $\ell \to \infty$  limit of our results above.  After doing so, we may remove the cutoff by taking the further limit $\rho_D \to \infty$.  We obtain the desired results by rewriting the generators \eqref{xi diff} in the form
\begin{equation}\label{bmn redef finite rho}
	{\cal P}_m = \frac{1}{\ell}(\chi_m + \bar \chi_{-m}), \qquad {\cal J}_m = \chi_m - \bar \chi_{-m}.
\end{equation}
Using \eqref{ip rhoD} we find
\begin{equation}\label{ip bmn rhoD}
	({\cal L}_{P_m} g, {\cal L}_{J_n} g) = \frac{1}{4 G} \frac{m(8 M G + m^2)}{f(\rho_D)} \hat L _1(\rho_D) \hat L_2(\rho_D)^*
	 \delta_{m, n},
\end{equation}
\noindent where we have rescaled $L(\rho) = \hat L (\rho) \ell^2 /2 $. Choosing $\hat L_{1,2}(\rho_D)$ independent of $\ell$ gives a finite result in the limit $\ell \to \infty$. Due to  \eqref{L asympt} we require $\hat L_{1,2}(\rho_D) = \rho_D + \ldots$, so finally taking $\rho_D \to \infty$ yields
\begin{equation}\label{ip bmn rhoD}
	({\cal L}_{P_m} g, {\cal L}_{J_n} g) = \frac{1}{4 G} m(8 MG + m^2) \delta_{m, n}.
\end{equation}
\noindent This coincides with the 2+1 BMN algebra \cite{Barnich:2006av} evaluated on a spacetime of energy $M$.

\section{Discussion}
\label{disc}

The above work considered Einstein-Hilbert gravity in $d$ spacetime dimensions with Dirichlet walls at finite distance; i.e., with a finite cutoff.  We examined physical excitations described by linearized diffeomophsisms and their relation to asymptotic symmetries that arise when the cutoff is removed.  In particular, for $d \ge 4$ with zero or negative cosmological constant we were able to see the emergence of the AdS and Poincar\'e groups, and for $d=3$ we obtained the full AdS$_3$ double-Virasoro algebra and correspondingly infinite 2+1 BMS group.

While we did not complete the task of carefully constructing the infinite volume phase space as a limit -- an in particular of proving from the results at finite $\rho_D$ that the approximate symmetries become exact as $\rho_D \rightarrow \infty$ -- it seems clear that this can be done.
An interesting general question in this context is the extent to which approximate symmetries of the regulated system may continue to be described as pure diffeomorphisms when acting on truly general solutions (e.g., which might contain matter near the Dirichlet wall) or at higher orders.  But for pure 2+1 Einstein Hilbert gravity, the formulation of these perturbations in section 5 of \cite{Marolf:2012dr} does indeed extend to define finite amplitude diffeomorphisms at fixed $\rho_D$ -- at least at the level of counting degrees of freedom, meaning that it leads to a single partial differential equation for a single function.

For $d \ge 4$ the AdS and Poincar\'e asymptotic symmetries generate rotationally invariant $(j=0)$ or dipole perturbations $(j=1)$.  But we have also looked for $j > 1$ diffeomorphisms of Schwarzschild and Schwarzschild-AdS$_d$ which preserve Dirichlet boundary conditions on a cylinder $S^{d-2} \times \mathbb{R}$.  These do not exist.  So the precise method used to study the 2+1 BMS group above does not yield the BMS group in higher dimensions.  However, in parallel with the finite amplitude comments above, it may be that one can obtain useful insight into how a higher-dimensional BMS group might act on a gravitational phase space by considering a larger set of perturbations in the regulated Dirichlet wall system.  This would provide a new implementation of the idea \cite{Strominger:2013jfa,Kapec:2014opa,Strominger:2014pwa,Pasterski:2015tva,Kapec:2015vwa} that BMS transformation are the soft (i.e., long-wavelength) limit of gravitons.  For example, it may be instructive to consider the lowest normal mode for each angular momentum $j$ and to find some sense in which these approach pure diffeomorphisms when the distance to the wall is taken to infinity.

\acknowledgments

It is a pleasure to thank Ted Jacobson for discussions that led to this project as well as encouragement to write up the results.   We also thank Will Kelly for collaboration on and discussions surrounding the closely related work \cite{Andrade:2015gja} and Andr\'es Anabal\'on 
for helpful comments. 
T.A. was supported by the European Research Council under the European Union's Seventh Framework Programme
(ERC Grant agreement 307955).
He also thanks the Centro de Ciencias de Benasque Pedro Pascual for their hospitality during the completion of this work.
D.M. was supported by the National Science Foundation under grant numbers PHY12-05500 and PHY15-04541 and by
funds from the University of California. He also thanks the KITP for their hospitality during the initial stages of the project,
where his work was further supported in part by National Science foundation grant number PHY11-25915.

\appendix

\section{Symplectic structures in a box}
\label{SSB}

We now briefly review the discussion from \cite{Andrade:2015gja} of the symplectic structure for theories with Dirichlet walls.  The symplectic current receives
a contribution from the Gibbons-Hawking term which plays a crucial role in its conservation -- unless one works in radial gauge where this contribution vanishes.

Before addressing the details of the gravitational system we briefly summarize the general procedure \cite{Compere:2008us} for constructing a conserved symplectic structure from a well-defined variational principle for a field theory in the presence of a boundary.  See also \cite{Lee:1990nz,Iyer:1994ys,Wald:1999wa} for related treatments of covariant phase spaces which do not study such boundaries in detail.  We denote the (not necessariy scalar) fields by $\phi$ and assume that the action
\begin{equation}
	S[\phi] = \int_M L_0 + \int_{\partial M} L_{\partial}
\end{equation}
\noindent has an extremum for some boundary condition $b(\phi) =0$.  This $b$ can be any local functional of the fields $\phi$.  Thus
\begin{equation}\label{Dbc condition app}
	\delta S = \int_{\partial M} \pi_b \delta b
\end{equation}
\noindent when the bulk equations of motion hold, and this $\pi_b$ may be called the momentum conjugate to $b$. As usual, we take $\partial M$ to be the part of the boundary where boundary conditions need to be imposed in order to define a phase space. In particular, we neglect any terms lying at past or future boundaries of the system.

Varying the bulk term yields
\begin{equation}
	\delta L_0 = ({\rm eoms}) \delta \phi + d \theta_0,
\end{equation}
\noindent and \eqref{Dbc condition app} implies that the pull-back of $\theta_0$ to $\partial M$ satisfies
\begin{equation}
\label{eq:sympkey}
	\theta_0|_{\partial M} = \pi_b \delta b - \delta L_\partial + d \theta_\partial
\end{equation}
for some $\theta_\partial$.  The total derivative $d \theta_\partial$ does not contribute to \eqref{Dbc condition app} since we again neglect terms lying at any past or future boundaries.

Following \cite{Compere:2008us}, we take the symplectic current to be
\begin{equation}
 	j = j_0 - d j_\partial,
 \end{equation}
\noindent where $j_0$ and $j_\partial$ are the symplectic currents associated to the potentials
$\theta_0$ and $\theta_\partial$, i.e.
\begin{equation}
	j_0 = \delta_2 \theta_0[\delta_1 \phi] - \delta_1 \theta_0[\delta_2 \phi], \qquad
	j_\partial = \delta_2 \theta_\partial[\delta_1 \phi] - \delta_1 \theta_\partial[\delta_2 \phi].
\end{equation}
Since the anti-symmetric second variation of $L_\partial$ vanishes identically, the anti-symmeric variation of \eqref{eq:sympkey} requires $j$ to vanish when pulled back to $\partial M$ ($j|_{\partial M} =0$) and evaluated on variations satisfying the desired boundary condition (so that $\delta b=0$).  There is thus no flux of symplectic current though the boundary, and conservation of the symplectic structure $\int_{\Sigma} j$ follows immediately from the fact that the bulk contribution to this current is closed ($d j_0 = 0$, see \cite{Lee:1990nz,Iyer:1994ys,Wald:1999wa}) so long as the hypersurface $\Sigma$ has boundaries only on $\partial M$.

We wish to follow the above procedure for gravity with Dirichlet boundary conditions. We consider
space-times for which $\partial M$ is a time-like surface of constant radial coordinate $r$,
with unit normal $n_\mu$.
We assume our spacetimes can be foliated near $\partial M$ by constant $r$ surfaces, on which we introduce coordinates $y^i$,
so that the metric can be written in the form
\begin{equation}
	ds^2 = N^2 dr^2 + \gamma_{ij} (dy^i + N^i dr) (dy^j + N^j dr),
\end{equation}
\noindent where the induced metric on surfaces of constant $r$ is $\gamma_{ij}$ and $N$, $N^i$ are the radial lapse and shift
functions, respectively.
Note that the normal satisfies $n_\mu dx^\mu = N dr$.

Since we impose Dirichlet boundary conditions on $\partial M$, the Einstein-Hilbert action with Gibbons-Hawking boundary term provides a valid variational principle \cite{Wald:1984rg}:
\begin{equation}\label{total action app}
 	S = \frac{1}{16\pi G} \int_M \sqrt{g} (R - 2 \Lambda) + \frac{1}{8\pi G}  \int_{\partial M} \sqrt{\gamma} K,
\end{equation}
\noindent where $\gamma_{\mu \nu} = g_{\mu \nu} - n_\mu n_\nu$ is the induced metric at the boundary
and $K$ is the trace of the extrinsic curvature $K_{\mu \nu} = \gamma_\mu \hs^\sigma \nabla_\sigma n_\nu$. 
In this covariant notation, tensors on $\partial M$ are degenerate space-time tensors which vanish when contracted with $n_\mu$.
In particular, we have $\gamma = {\rm det} \, \gamma_{ij}$ and $\gamma \neq {\rm det} \, \gamma_{\mu \nu} = 0$. \\

The bulk contribution to the symplectic current is the standard one for Einstein-Hilbert gravity, which as in \cite{Andrade:2009ae} we take to be given by%
\begin{equation}\label{j EH}
	j^\nu_{EH} = \frac{1}{16\pi G} [ \delta_2 ( \sqrt{g} g^{\alpha \beta} ) \delta_1 \Gamma^\nu _{\alpha \beta} - \delta_2 ( \sqrt{g} g^{\alpha \nu} ) \delta_1 \Gamma^\beta_{\alpha \beta}
	- (1 \leftrightarrow 2) ].
\end{equation}
See \cite{Iyer:1994ys,Barnich:2001jy, Barnich:2007bf} for other choices of symplectic currents that differ from \eqref{j EH} by total derivatives.
A general on-shell variation of the action \eqref{total action app} is of the form
\begin{equation}
	\delta S = \int_{\partial M} \sqrt{\gamma} ( \pi^{\mu \nu} \delta \gamma_{\mu \nu} + {\cal D}_\mu c^\mu),
\end{equation}
\noindent where ${\cal D}_\mu$ is the covariant derivative compatible with $\gamma_{\mu \nu}$, the conjugate momentum is given by
\begin{equation}
	\pi^{\mu \nu} = - \frac{1}{16\pi G}  ( K^{\mu \nu} - K \gamma^{\mu \nu} ),
\end{equation}
\noindent and
\begin{equation}
 	c^\mu = - \gamma^{\mu \rho} \delta g_{\rho \sigma} n^\sigma
 \end{equation}
is tangent to $\partial M$ so that ${\cal D}_\mu c^\mu$ is well-defined.

The Dirichlet condition sets $\delta \gamma_{\mu \nu} |_{\partial M} = 0$ so that the boundary contribution to the symplectic potential becomes
\begin{equation}
	\theta_\partial^i = - \frac{1}{16\pi G}  \sqrt{\gamma} g^{i \lambda}  \delta g_{\lambda \sigma} n^\sigma.
\end{equation}
The antisymmetrized variation then yields
\begin{equation}
\label{bndycurrent}
	j_\partial^i = - \frac{1}{16\pi G}  [ \delta_2 (\sqrt{\gamma} g^{i \lambda} n^\sigma) \delta_1 g_{\lambda \sigma} + (1 \leftrightarrow 2)],
\end{equation}
and the total symplectic structure is
\begin{equation}\label{Omega full}
	\Omega = \int_\Sigma j_{EH} - \int_{\partial \Sigma} j_\partial .
\end{equation}

\bibliographystyle{JHEP-2}

\bibliography{BoxBib}

\end{document}